\documentclass[12pt, preprint,numberedappendix]{emulateapj}

\newcommand\submitms{n}		
\newcommand\bibinc{n}		
\tabletypesize{\scriptsize}

\newcommand*\wrapletters[1]{\wr@pletters#1\@nil}
\def\wr@pletters#1#2\@nil{#1\allowbreak\if&#2&\else\wr@pletters#2\@nil\fi}

\usepackage{subeqnarray}
\usepackage{natbib}
\usepackage{tweaklist}
\renewcommand{\itemhook}{\setlength{\topsep}{1pt}%
  \setlength{\itemsep}{2pt} \setlength{\parskip}{0pt} \setlength{\parsep}{0pt}}

\newcommand{\ie}{i.e.\ }
\newcommand{\eg}{e.g.\ }
\newcommand{\p}{\partial}
\newcommand{\xv}{\vc{x}}
\newcommand{\zv}{\vc{z}}

\newcommand{\brak}[1]{\langle #1\rangle}

\newcommand{\gcc}{\;\mathrm{g/cm^{3}}}

\def\K{\; {\rm K}}

\newcommand{\vc}[1]{\mbox{\boldmath{$#1$}}}

\DeclareMathSymbol{\varOmega}{\mathord}{letters}{"0A}
\DeclareMathSymbol{\varSigma}{\mathord}{letters}{"06}
\DeclareMathSymbol{\varPsi}{\mathord}{letters}{"09}

\newcommand{\Eq}[1]{Equation\,(\ref{#1})}
\newcommand{\Eqs}[2]{Equations (\ref{#1}) and~(\ref{#2})}

\newcommand{\Fig}[1]{Fig.~\ref{#1}}
\newcommand{\Figs}[2]{Figs.~\ref{#1} and \ref{#2}}
 
\newcommand{\Tab}[1]{Table \ref{#1}}

\def\Nexp{N_{\rm exp}}
\def\Npl{N_{\rm pl}}
\def\Keplerp{\emph{Kepler}} 
\def\Kepler{\emph{Kepler} } 
\def\days{~{\rm days}}
\def\dxN{d\xv}
\def\pxN{\p\xv^{N_x}}
\def\av{\vc{\alpha}}

\def\etadisc{\eta_{\rm disc}}
\def\mF{\mathcal{F}}

\slugcomment{Draft Modified \today}

\shorttitle{\emph{Kepler}'s Exoplanet Census}
\shortauthors{Youdin }

\begin{document}

\title{The Exoplanet Census: A General Method, Applied to \emph{Kepler}}
\author{Andrew N.\ Youdin}
\affil{Harvard Smithsonian Center for Astrophysics, 60 Garden Street, MS-16, Cambridge, MA 02138, USA}

\begin{abstract}
We develop a general method to fit the planetary distribution function (PLDF) to exoplanet survey data.  This maximum likelihood method accommodates  more than one planet per star and any number of planet or target star properties.  Application to \Kepler data relies on estimates of the efficiency of discovering transits around Solar type stars by Howard et al. (2011).  These estimates are shown to agree with theoretical predictions for an ideal transit survey.  Using announced \Kepler planet candidates, we fit the PLDF as a joint powerlaw in planet radius, down to $0.5 R_\oplus$, and orbital period, up to 50 days.   The estimated number of planets per star in this sample is $\sim 0.7$ ---$1.4$, where the broad range covers systematic uncertainties in the detection efficiency.  To analyze trends in the PLDF we consider four planet samples, divided between shorter and longer periods at 7 days and between large and small radii at 3 $R_\oplus$.      At longer periods, the size distribution of the small planets, with index $\alpha \simeq -1.2\pm 0.2$ steepens to   $\alpha \simeq -2.0\pm0.2$ for the larger planet sample.  For shorter periods, the opposite is seen: smaller planets follow a steep powerlaw, $\alpha \simeq -1.9\pm0.2$ that is much shallower, $\alpha \simeq -0.7\pm0.2$ at large radii.
The observed deficit of intermediate-sized planets at the shortest periods may arise from the evaporation and sublimation of Neptune and Saturn-like planets.  If the trend and explanation hold, it would be spectacular observational confirmation of the core accretion and migration hypotheses, and allow refinement of these theories.
\end{abstract}

\section{Introduction}
Individual exoplanet discoveries highlight the extraordinary diversity of worlds in the Solar neighborhood \citep{MQ95,70vir,GJ1214,Kep11}.     For an accurate census of the planet population, the statistical analysis of large samples of exoplanets is required \citep{cumming08,howard2010}.  Trends revealed by the data provide powerful constraints on dynamical theories of planet formation \citep{gls04,kb06,houches10}.

The correlation of giant planets with host star metallicity is perhaps the most interesting trend revealed by radial velocity surveys \citep{gonzalez1997,fv05,john10}.  This trend has been a  powerful guide to identifying the mechanisms responsible for planetesimal formation \citep{ys02,yg05,jym09}, see \citet{cy10} for a review.

The \Kepler transit survey is currently revolutionizing the field of exolanets from space \citep{KepMission}.   \citet[hereafter BKep11]{borucki_pc11} released 1,235 partially vetted planet ``candidates."  \citet{mj2011} estimate the false positive rate as $\lesssim 5\%$ around the brighter stars that are considered here.  For brevity and statistical purposes, we mostly refer to the candidates as ``planets," though significant followup work remains with only 18 candidates currently confirmed.

\citet[hereafter HKep11]{Howard2011} presented a detailed statistical analysis of the $\sim 500$  BKep11 planets around $\sim 60,000$ bright, Solar-type stars.  Accounting for detection efficiencies, HKep11 report planet occurrence rates of $\sim 0.2$ planets per star for radii $> 2 R_\oplus$ and orbital periods $< 50 \days$.  The purpose of this paper is to apply different statistical methods to the \Kepler dataset, making use of the estimates of detection efficiencies --- whose importance is on par with the detections themselves --- provided by HKep11.  

Our method is based on the \citet[hereafter TT02]{tt02} technique to analyze the mass and period distributions of planets from radial velocity surveys.  
TT02 emphasized the importance of determining  planet mass and period distributions simultaneously.  A simultaneous fit is crucial because radial velocity detection thresholds depend on both the mass (times the sine of inclination) and period.  Transit surveys have a similar, though weaker, coupling between planet radius and orbital period in the efficiency of discovering transits.

A primary advantage of the TT02 technique is that it naturally accommodates more that one planet per star.  Planet hosting is not treated as a binary proposition.  This distinction is not just a technical nicety, because the number of planets per star is order unity, at least.  We show that \Kepler data already imply about one planet per Solar type star within $\sim 0.25$ AU and with a radius bigger than  $0.5 R_\oplus$.  This number is certain to grow with time as longer periods and smaller sizes are probed.

High multiplicity rates in the \Kepler sample \citep{latham_mult11}  means that the fraction of stars that host a planetary system (at most one, of course) can be significantly smaller than the number of planets per star (NPPS).      We treat the planet population (including known multiples) as a whole, and ignore multiplicity issues.   \cite{rh10} discuss the value of multi-transiting systems in constraining mutual inclinations.  Using these constraints, \Kepler data show that the number of planets per planetary system   is at least 2 --- 3 \citep{lr11}.

This paper is organized as follows.  Section \ref{sec:selection} describes the detection efficiencies relevant to the \Kepler transit survey, including an analytic fit to  discovery efficiencies reported by HKep11 in \S\ref{sec:kepeff}.  Section \ref{sec:method} describes our method for analyzing exoplanet survey data.  Section \ref{sec:results} applies the method by fitting \Kepler data to a joint powerlaw in radius and period.  Differences between the shorter and longer period planets (in \S\ref{sec:shortlong}) and also between the smaller and larger planets (in \S\ref{sec:quad}) are analyzed.  Section \ref{sec:npps} gives non-parametric estimates of planet occurrence, and comments speculatively on the consequences of rising period distributions.  We conclude with a summary and  discussion in \S\ref{sec:disc}.  Appendix \ref{sec:analapp} gives an analytic solution for powerlaw fits to data from an idealized transit survey.

\begin{figure*}[tb!] 
\if\submitms y
 	\includegraphics[width=3.3in]{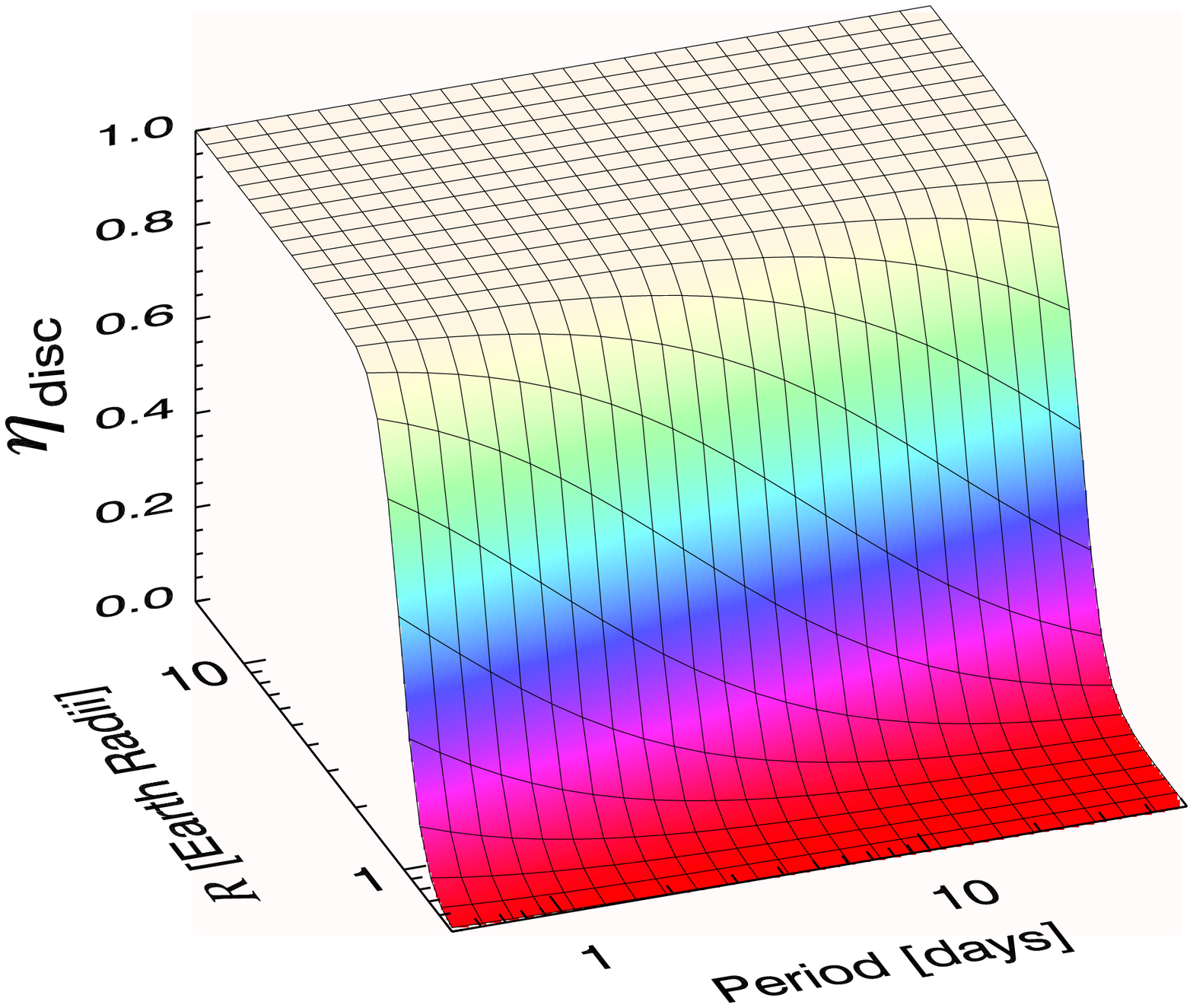}
	\includegraphics[width=3.3in]{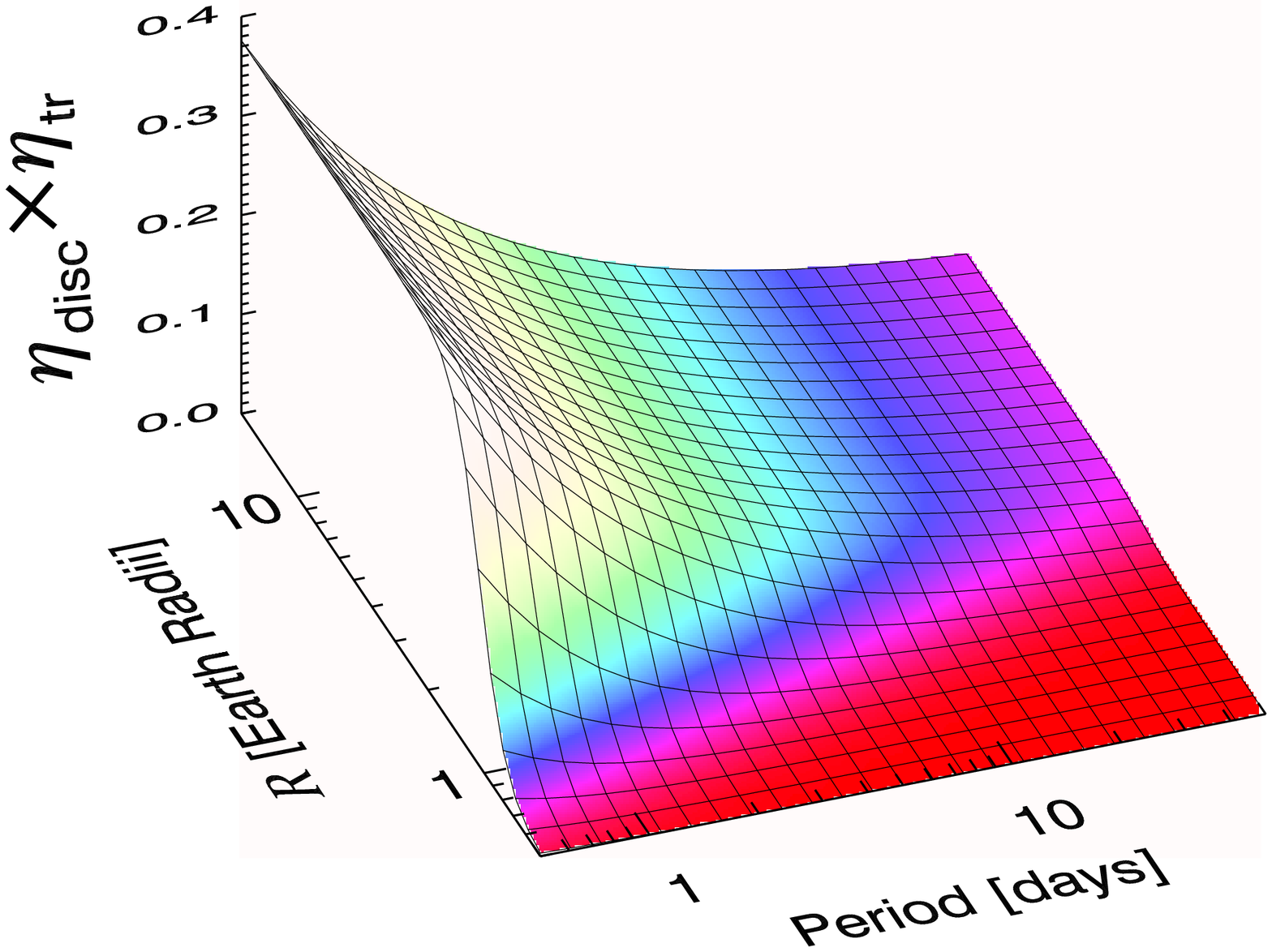}
 \else
	\hspace{-1cm}
          \includegraphics[width=3.9in]{shadeyx.eps} 
            \includegraphics[width=3.9in]{shadenet.eps} 
\fi
   	\caption{(\emph{Left}): The  \Kepler mission's efficiency of  discovering transiting planets around bright Solar-type stars vs.\ planetary radius and orbital period. 
	The discovery efficiency is unity for large planets, but drops sharply below a planet size that increases gradually with orbital period.  See text for details. (\emph{Right}): The net detection efficiency combines the  discovery efficiency 
	(at left) with the geometric transit probability, which exerts a bias against detecting long period planets of all sizes.}
   	\label{fig:deteff}
\end{figure*}

\begin{figure*}[tb] 
\if\submitms y
 	\includegraphics[width=3.3in]{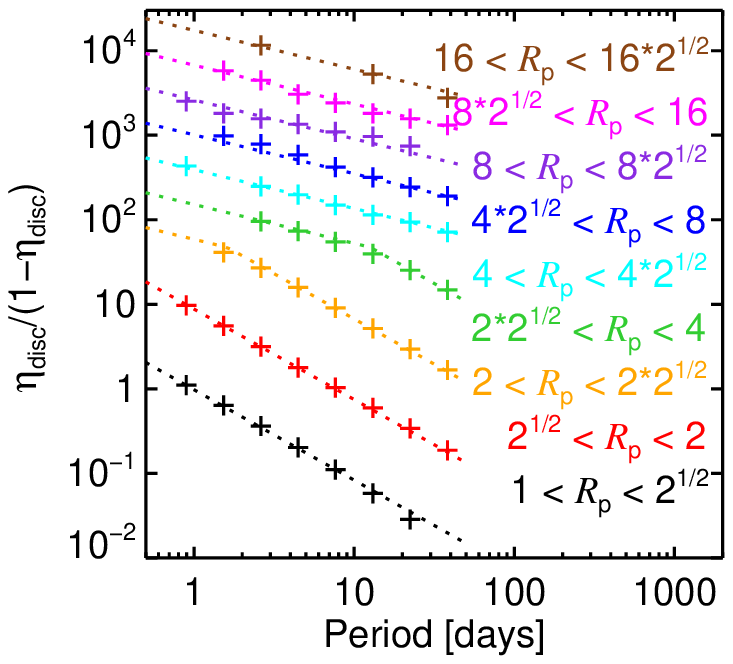}
 	\includegraphics[width=3.3in]{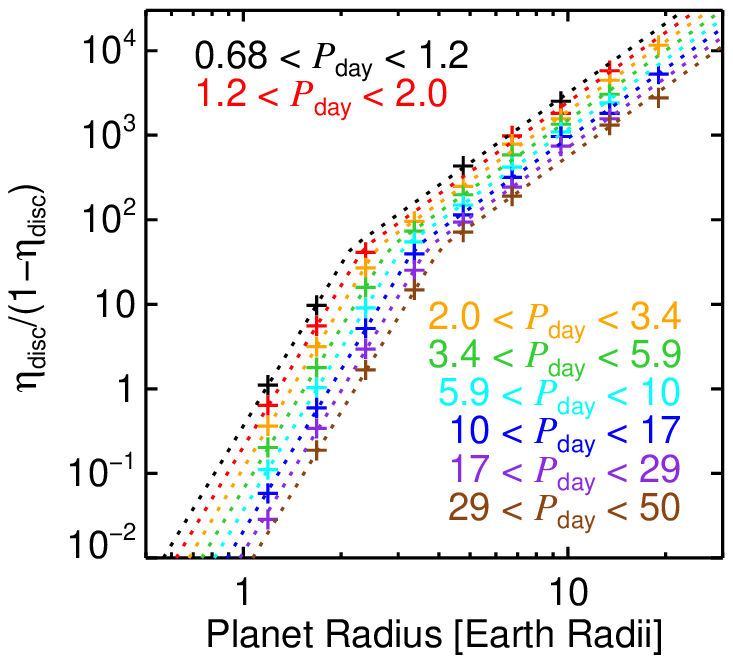}
 \else
	\hspace{-1cm}
          \includegraphics[width=3.9in]{RatioVsPer.eps} 
          \hspace{-.5cm}
            \includegraphics[width=3.9in]{RatioVsRad.eps} 
\fi
   	\caption{The \Kepler discovery efficiency plotted as a ratio, $\etadisc/(1-\etadisc)$ of detectable-to-non-detectable planets.  Symbols correspond to the values reported in Figure 4 of HKep11, while the curves give the broken powerlaw fit of \Eqs{eq:discratio}{eq:discratiob}.  At left, the ratio is plotted vs.\  orbital period for a range of (binned) planetary radii.  The same data are plotted vs.\ planet radius at right.  The smooth behavior of the \Kepler discovery efficiencies is  evident.} 
   	\label{fig:discrat}
\end{figure*}

\section{Selection Effects For Transit Surveys}\label{sec:selection}
A robust statistical analysis must account for relevant selection effects.  We quantify selection effects as detection efficiencies, $\eta$,  which give the ratio of detections to actual planets.  
Individual efficiencies multiply to give the net detection efficiency.  

The three main selection effects for transit surveys are:  (i) the transit probability for a planetary orbit to cross our line of sight to the star, $\eta_{\rm tr} \leq 1$; (ii) the discovery efficiency of the survey in finding transiting planets, $\etadisc \leq 1$; and (iii) the rate of false positive events that mimic a planet transit, $r_{\rm fp} \leq 1$.  For false positives, the relevant  $\eta_{\rm fp} = 1/(1-r_{\rm rp}) \geq 1$, the only efficiency that can exceed unity.   As mentioned already, the false positive rate is estimated to be low enough that we ignore it for simplicity.

For most analyses, an average detection efficiency is insufficient.  The dependence of the efficiencies on relevant properties of planets and target stars is required.  We consider how the efficiencies vary with planet radius, $R$, and orbital period, $P$.  See \S\ref{sec:whichparam} for a general discussion of which parameters should be included.

The left panel of \Fig{fig:deteff} shows the \Kepler discovery efficiency, calculated as described below (in section \ref{sec:kepeff}).  The discovery efficiency is nearly unity over a signifiant range of planetary radii and periods, thanks to the high photometric precision of \Keplerp.  There is a sharp drop in discovery efficiency for sufficiently small planets whose transit depths compete with noise in the photometric data.  At shorter periods, the signal from smaller  planets can rise above the noise because more transits are seen.  The right panel of \Fig{fig:deteff} shows how the geometic transit probability reduces the probability of finding long period planets.

To connect with the extensive work on radial velocity (RV) surveys, we note that RV upper limits can be expressed as a detection efficiency for a  given star.  Most simply, $\eta = 1$ for planet parameters above the detection threshold and $\eta = 0$ otherwise.  This step function can be smoothed to account for confidence intervals.

\subsection{Transit Probability}
The probability that a planet, on a randomly oriented circular orbit of semi-major axis $a$, transits it's host star, with radius $R_\ast$ and mean density $\rho_\ast$, is
\begin{eqnarray} \label{eq:etatr}
	\eta_{\rm tr} &=& {R_\ast \over a} 
	= \left(3 \pi\over  G \rho_\ast  P^{2}  \right)^{1/3} \\
	&=& 0.051 \left(10 \days \over P \right)^{2/3} \left(\rho_\odot \over \rho_\ast \right)^{1/3} \nonumber
\end{eqnarray} 
for $R \ll R_\ast$.
Eccentricity tends to increase the detectability of planets at fixed $a$ \citep{burke08}.\footnote{For an instructive explanation see \wrapletters{\tt http://oklo.org/2011/03/23/the-eccentricity-distribution/}}     For the mean eccentricities 
$\sim 0.1$ --- $0.25$ implied by \Kepler transit durations \citep{moorhead11}, corrections are $\lesssim 10\%$ and ignored here.

For simplicity, we approximate the mean stellar density (or more specifically the mean $\rho_\ast^{-1/3}$) as Solar.  The densities of planet hosts can be estimated from precise light curve parameters, though corrections for the eccentricity apply \citep{tingley11}.   
Furthermore, planet hosts have a lower average density than the target star population, on general.  This bias arises because lower density stars have higher transit probabilities.  For our purposes, uncertainties in the average stellar density modestly affects the magnitude (but not the shape) of the inferred planet distribution.

One might (incorrectly) expect multi-planet systems to require special values of  $\eta_{\rm tr}$.  When mutual orbital inclinations are small, finding one planet does increase the odds that others will be found.  
However since the planetary systems as a whole are randomly oriented, low mutual inclinations also make it easier to miss an entire planetary system.
The mutual inclination of planets within systems has no effect on the overall detection rate (aside from sampling noise, as always).

\subsection{The {\it Kepler} Discovery Efficiency}\label{sec:kepeff}
Precisely quantifying the  discovery efficiency of a survey is quite difficult and is ongoing work for the \Kepler mission.  For our study we rely on the estimates of discovery efficiency in HKep11.  HKep11 quantified the discovery efficiency for a subsample of bright solar-type stars with relatively high $\etadisc$.  The sample consists of $N_\ast = 58,041$ stars that satisfy the effective temperature, surface gravity and \Kepler magnitude cuts:
\begin{subeqnarray}\label{eq:starcuts}
& 4100\K \leq T_{\rm eff} \leq 6100\K&  \\
&4.0 \leq \log g /({\rm cm}~{\rm s}^{-2}) \leq 4.9& \, \\
&{\rm K_P} \leq 15 ~{\rm mag}&\, .
\end{subeqnarray} 
Section \ref{sec:planetsample} discusses the planet candidates in this sample. 

The  data for $\etadisc$ is presented in Figure 4 of HKep11.  Here the discovery efficiency is reported for planets with $ 0.68 <  P/({\rm days}) < 50 $ and $1 < R/R_\oplus < 16$ on a log-uniform grid.  HKep11 report $\etadisc N_\ast$, the number of stars around which a planet with that radius and period could be detected, in the bottom left of each cell.  The related values of $\etadisc$ are plotted in \Fig{fig:discrat}.

For bins with no detected planets, HKep11 do not report a discovery efficiency.  This omission is not because planet detections are required to estimate efficiencies.  Rather HKep11  applied efficiencies to the detections only, which differs from our approach of applying the efficiencies across all parameter space.  The missing efficiencies in Figure 4 of HKep11 can be readily obtained by interpolation.  Most of the empty bins correspond to small $P$ and large $R$ where $\etadisc \approx 1$.  Thus the lack of detections in these bins carries heavy statistical significance.

Instead of merely interpolating the reported  \Kepler efficiencies, we are motivated by their smooth variation to find an analytic fit.  We find an excellent joint powerlaw fit, not to $\etadisc$ itself, but to the related
\begin{equation} 
r_{\rm disc}\equiv {\etadisc \over 1 - \etadisc} \, , 
\end{equation} 
 the ratio of discoverable to non-discoverable planets. Since $r_{\rm disc}$ ranges from zero to arbitrarily large values, it is simpler to fit with a powerlaw.  The restriction to $\etadisc \leq 1$ is automatically satisfied.

\Fig{fig:discrat} shows that the broken powerlaw is an excellent fit to reported discovery efficiencies.  
The powerlaw fit is
 \begin{equation} \label{eq:discratio}
	 r_{\rm disc} = 0.324 \left(R \over R_\oplus\right)^{6.34} \left(P \over {\rm day} \right)^{-1.07}
 \end{equation}
 below a break at 
 \begin{equation} 
	R < R_{\rm b} = 2.2 \left(P \over {\rm day}\right)^{0.17} R_\oplus \, .
\end{equation}
The efficiency at the break is quite high, $\etadisc = 0.98$ (and remarkably constant, probably reflecting the high S/N threshold).  Note that $\etadisc$ drops to 50\% ($r_{\rm disc} = 1$ in \Fig{fig:discrat}) at $R = 0.54 R_{\rm b}$, rising roughly from 1 to 2 $R_\oplus$ over the periods considered.  When discovery efficiencies are small, $\etadisc \approx r_{\rm disc} \ll 1$,  the powerlaw in \Eq{eq:discratio} applies directly to $\etadisc$.

While the scalings for  $\etadisc > 0.98$ are of little practical concern for our study,  we report the fit above the break for completeness:
\begin{equation}\label{eq:discratiob} 
r_{\rm disc}(R > R_{\rm b}) = 5.46\left(R \over R_\oplus\right)^{2.74} \left(P \over {\rm day} \right)^{-0.460} \, .
\end{equation}
This fit describes a small fraction of noisy stars, but does not affect our results because $\etadisc \approx 1$ in this regime.
  
The relevant fit in \Eq{eq:discratio} agrees well with theoretical estimates.  For  S/N limited detections, the combined efficiency should scale as 
\begin{equation} 
\etadisc \eta_{\rm tr} \propto P^{-5/3} R^6\, ,
\end{equation} 
see \eg Equation (7) of \citet{gaudi07}.  The small $R$ limit of \Eq{eq:discratio} gives $\etadisc \eta_{\rm tr} \propto P^{-1.74} R^{6.34}$, rather good agreement.

The efficiencies summarized here will likely be improved by more detailed studies that include extraction of simulated transits from the actual \Kepler data analysis pipeline.  Accurate estimates of the \Kepler discovery efficiency for all relevant parameters --- especially stellar $T_{\rm eff}$ --- are the most crucial ingredient for determining the frequency of Earth-like planets.

\subsection{The Planetary Sample}\label{sec:planetsample}
Tables 1 and 2 of BKep11 list the properties of 1,235 planet candidates and their host stars.  We must restrict our attention to the planet candidates around host stars with known discovery efficiencies, $\etadisc$.  Thus we only consider host stars within the Solar sample of HKep11, set by \Eq{eq:starcuts}.  We further restrict attention to detections with signal-to-noise (S/N) $>$ 10 and $P < 50$ days, as these are also conditions for the validity of $\etadisc$.  Applying these filters reduces the number of planet candidates from 1,235 to 566.  

Our ``full" planet sample considers 
\begin{subeqnarray}
 0.5 \days < P < 50 \days \\
 0.5 ~ R_\oplus <  R <  20 ~R_\oplus \, .
\end{subeqnarray} 
Adding these radius and minimum period cuts only removes four planets (from 566 to 562).  Instead of using the BKep11 reported values for $R$, we compute $R$ from the reported values for the host star radius, $R_\ast$, and $R/R_\ast$.\footnote{This step overcomes the rounding of the reported $R$ to $0.1 R_\oplus$ intervals, and is merely for convenience in defining boundaries in an assumption-free way.}

We now discuss relevant differences between our planet sample and that of HKep11.  HKep11 conservatively restricted attention to $R > 2 R_\oplus$, because the discovery efficiency is high for these planets (except at longer periods).  We extend our analysis down to $R = 0.5 R_\oplus$, which includes essentially all small planets in the Solar sample.  This extension involves (a) trusting the HKep11 reported efficiencies down to the $1 R_\oplus$ level to which they were reported, and (b) extrapolating the efficiencies down to $0.5 R_\oplus$.   As shown above, both the smooth variation of the reported efficiencies and their agreement with theory give us confidence in making the extrapolation.  We also report results for samples of larger planets that are unaffected by this extrapolation.

HKep11 defined S/N slightly differently than the publicly reported BKep11 S/N values on which we rely.  HKep11 define S/N based on only one quarter of data (Q2,  \Keplerp's first full quarter).  By contrast BKep11 report S/N values up to Q5, so that roughly four times more data has been collected.  With perfect noise scaling,  the BKep11 S/N values should be roughly twice as high as those used by HKep11.    Table 2 of HKep11 presents noise scalings for 4 planets with $\sim 2.5 R_\oplus$ radii and $\sim 40$ day periods.  The ideal scaling roughly holds for three objects, but S/N saturated and only rose modestly for the fourth.    It's unclear what the general noise scaling is, and how it varies with size and period.

We proceed conservatively by performing all our analyses on both a S/N $> 10$ and a S/N $> 20$ planet sample (as defined by BKep 11).  These cases cover the complete range of possibilities between saturated noise and perfect scaling.  Fortunately the adopted S/N threshold has a modest effect on quantitative results (most importantly, the number of planets per star) but no effect on the interesting trends we discuss.

We briefly note a discrepancy between the planet counts reported by BKep11 and HKep11.  HKep11 report a sample of 438 planet candidates (of the 1,235 released by BKep11) that  (a) have hosts that satisfy the stellar parameter cuts of \Eq{eq:starcuts}, (b) have planet parameters $P \leq 50$ days and $ 2 R_\oplus \le R \le 32 R_\oplus $ and (c) satisfy the HKep11 definition of S/N $>$ 10.  If we apply cuts (a) and (b) only to the BKep11 tables, but allow all S/N, there are 378 candidates.  That number increases to 393 using the reported $R$ instead of calculating it as $(R/R_\ast) \times R_\ast$.   Thus there are at least 60 (or 45) planets that HKep11 say should appear in the BKep11 tables, but they do not.

Applying any S/N threshold to the BKep11 tables can only make the discrepancy larger.  We made no choices that would cause us to miss planets in the BKep11 tables.  Cuts were applied inclusively (including any values that fall on a boundary) and all the planets in any system were counted.  We conclude that a discrepancy exists between the planet and/or stellar parameters used by HKep11 and reported by BKep11.  This discrepancy does not affect our results as long as the estimates of the discovery efficiencies in HKep11 are sufficiently accurate.    

\section{A General Method for Exoplanet Statistics}\label{sec:method}
This section describes a general method for estimating the planetary distribution function (PLDF) from the results of a survey with quantified detection efficiencies.  The purpose of developing this  formalism is twofold.

First, this  method can investigate a PLDF that depends on properties of both the planet detections and the target stars. Further, the PLDF can take an arbitrary functional form. While the exoplanet community is unlikely to agree on a single statistical methodology anytime soon, at least such an approach is possible.

Second, this formalism allows the inclusion of all detected planets --- including those in multi-planet systems.
This inclusion is possible because we treat the PLDF as the average number of planets per star (NPPS) not the fraction of stars with planets (FSWP).   Following TT02, our technique treats planet occurrence as a Poisson process, \ie a series of independent random events.   This technique naturally gives the NPPS.

By contrast, many statistical analyses treat planet hosting as a binary proposition, \ie a star either does or does not have a detected planet.  The use of binomial statistics  naturally gives the FSWP.   Including multiple planets per system in this type of analysis is formally incorrect, though discrepancies are small if the multiplicity rate is low.  For more discussion of these points see \citet[]{cumming08}, who rigorously analyze the FSWP in radial velocity surveys by only including the most detectable planet around any star.

For transit surveys,  the FSWP is more difficult to determine and is not addressed here.   The main issue is applying the transit efficiency correction.  Should the missed planets that do not transit be assigned evenly among all stars, or into high order multi-planet systems?   As mentioned in the introduction, multi-transiting systems are a powerful constraint, as are transit timing variations and comparison to RV surveys \citep{rh10}.  

Before developing the general method for fits to a parameterized PLDF in Section \ref{sec:method2}, we consider non-parametric estimates of the NPPS in Section \ref{sec:binfit}.
Parametric fits also give an estimate of the planet occurrence, whose quality depends on the appropriateness of the assumed functional form.  The main reason to consider parameterized fits is to identify and assess trends in the data --- and if feeling bold, to extrapolate.  Section \ref{sec:whichparam} discusses the consequences of fitting some parameters and ignoring others.

\subsection{Non-Parametric Estimates of Planet Occurrence}\label{sec:binfit}
For a survey of $N_\ast$ stars, we divide planet detections into bins indexed by $\ell$,  with $N_\ell$ planet detections in a given bin.  If the average detection efficiency per bin is $\eta_\ell$, then the best estimate of the NPPS in each bin is
\begin{equation} \label{eq:binfit}
f_\ell = {N_\ell \over \eta_\ell N_\ast}\, ,
\end{equation} 
the number of detected planets per star divided by the efficiency.

The total planet occurrence, $\sum_\ell f_\ell$, depends on bin size when $\eta_\ell$ is not constant.  For arbitrary small bins, there is only one planet per bin, and the efficiency is evaluated 
for the parameters of each planet and its host.   This small bin limit is the only truly assumption-free way to estimate the planet fraction, and suggests that binning of data is never required.

A maximum likelihood analysis using Poisson statistics can reproduce the estimates of $f_\ell$ given by \Eq{eq:binfit}, and develop intuition for the general method.  The number of expected detections in all bins is 
\begin{equation} 
\Nexp = N_\ast \sum_\ell \eta_\ell f_\ell\, .
\end{equation} 
The likelihood function for a Poisson process is
\begin{equation} \label{eq:likebins}
\tilde{L} = \left[ \prod_\ell(\eta_\ell f_\ell)^{N_\ell} \right] \exp \left(-\Nexp\right) \, ,
\end{equation} 
which represents the product of the probabilities of each individual detection (the term in square brackets) times the probability of finding no additional planets in any bin (the exponential). 

Any constant multiplicative factors can be ignored in the likelihood function, because they do not affect the maximization of likelihood.  (Thus factorials do not appear in \Eq{eq:likebins}.) Here constant means independent of the PLDF, \ie the $f_\ell$ values.  Remarkably, this freedom allows us to ignore the efficiencies that characterize the $N_\ell$ detections, but not in $\Nexp$.  The likelihood thus simplifies to
\begin{equation} \label{eq:likebinsimp}
L = \left[ \prod_\ell f_\ell^{N_\ell} \right] \exp \left(-\Nexp\right) \, .
\end{equation} 
The efficiencies now appear only in $\Nexp$.

The maximum likelihood values of $f_\ell$ are the roots of $\p L/ \p f_\ell = 0$, which reproduces the expected result of \Eq{eq:binfit}.  It is easier (analytically and numerically) to maximize $\ln(L)$. 

A fundamental feature of the method is that efficiencies are used to predict the number of expected planets and are not directly applied to the detections.\footnote{Non-detections however can enter $\Nexp$ via upper limits that define the efficiency, see \S\ref{sec:selection}.}  While this feature disappears when bin sizes are chosen to be arbitrarily small (as discussed above), it becomes important for parametric fits.

\subsection{Fitting the Planetary Distribution Function}\label{sec:method2}

We now describe a general method to fit a parameterized PLDF to unbinned data.   We consider a PLDF that depends on planet properties, $\xv$ (an $N_x$ component vector) and stellar properties $\zv$ (with $N_z$ components).  Usually the elements of these vectors  are logarithms of measured quantities such as the orbital period or stellar mass. 

The probability that a star with properties $\zv$ has a planet with properties $\xv$ that lie in a volume\footnote{For brevity the exponent in the phase space volume is dropped, \ie $d \xv \equiv d\xv^{N_x}$ here and throughout.  We keep this exponent in the denominator of derivatives to be explicit that these are not gradients, as in \Eq{eq:df}.} $d \xv$ of phase space is
\begin{equation} \label{eq:df}
df = {\p f (\xv,\zv) \over \pxN} \dxN \, .
\end{equation} 
Defined this way, the integrated $f$ represents the NPPS.  

We express the differential distribution,  
\begin{equation} \label{eq:dfdx}
{\p f (\xv,\zv) \over \pxN} \equiv C g(\xv,\zv ; \av)\, ,
\end{equation} 
as an amplitude, $C$, times a shape function, $g$.  The shape parameters, $\av$, can describe the behavior of individual planetary or stellar properties as well as any correlations.  In the simplest case, the values of $\av$  are powerlaw exponents.  

The total number of planets around $N_\ast$ stars  (indexed by $j$) is
\begin{subeqnarray} \label{eq:Ntot}
N_{\rm tot} &=& C \sum_j^{N_\ast} \int g(\xv,\zv_j;\av) \dxN \\
 &=& N_\ast C\int \mF_\ast(\zv) g(\xv,\zv;\av) \dxN d \zv  \slabel{eq:stellarint}\, .
\end{subeqnarray} 
where integration covers a specified volume of the phase space of $\xv$.  In \Eq{eq:stellarint} the sum over stars is replaced by an integral over the stellar distribution function $\mF_\ast(\zv)$.  We proceed with the integral notation to minimize indices, but direct summation over all stars is always possible (and preferred when feasible).

Consider a survey with a net detection efficiency
\begin{equation} 
\eta(\xv,\zv) = \prod_k \eta_k (\xv,\zv)
\end{equation} 
that is a product of efficiencies from individual selection effects (indexed by $k$).
The number of expected planet detections is
\begin{subeqnarray} \label{eq:Nexp}
\Nexp &=& N_\ast C \int  \eta (\xv,\zv) \mF_\ast(\zv) g(\xv,\zv;\av) d\xv d\zv\, , \\
&\equiv& N_\ast C F  \slabel{eq:F}
\end{subeqnarray} 
The shape integral, $F$, 
weights the shape function over both the stellar distribution and the efficiencies.

\subsubsection{Maximizing Likelihood}
Consider a survey that detects $\Npl$ planets around $N_\ast$ stars.  We define the likelihood of the data analogously to \Eq{eq:likebinsimp} as
\begin{equation} \label{eq:likebasic}
\tilde{L}= \left[ \prod_{i=1}^{\Npl} df_i \right] \exp \left(-\Nexp\right) \, ,
\end{equation}
which represents the probabilities of each individual detection, labelled by $i$, times the probability that no other planets were detected, given by the exponential factor.    

Applying \Eqs{eq:df}{eq:dfdx}, we again eliminate unnecessary constants, here the phase space volume,
to define the likelihood function as
\begin{eqnarray}
 L &=& \left[C^{\Npl} \prod_{i=1}^{\Npl} g(\xv_i, \zv_{j(i)};\av)\right] \exp \left(-\Nexp\right) \label{eq:Lg}\, ,
\end{eqnarray} 
where $j(i)$ refers to the star that hosts planet detection $i$.  We let $g_i$ represent the shape function evaluated for the properties of detection $i$ and express the log likelihood as
\begin{equation} \label{eq:logL}
\ln(L) = \Npl \ln (C) + \sum_{i = 1}^{\Npl} \ln(g_i) - \Nexp \, .
\end{equation} 

The best fit normalization, $C$, is found by maximizing the likelihood as the root of $\p \ln L/\p C = 0$:
\begin{equation} \label{eq:C}
C = {\Npl \over N_\ast F}
\end{equation} 
using \Eq{eq:F}.  

Following TT02, we use this constraint to eliminate $C$ from the likelihood:
\begin{eqnarray} \label{eq:LnoC}
\ln(L) &=& -\Npl \ln(F) + \sum_i^{\Npl} \ln(g_i) \\
&&+ \left\{\Npl \left[\ln \left(\Npl \over N_\ast\right)-1 \right]\right\} \nonumber
\end{eqnarray} 
where the constant term in curly brackets can again be ignored.

The best fit values of $\av$ maximize $L$ as the roots of $\p L /\p \av = 0$, which are
\begin{equation} \label{eq:alphamax}
{\p \ln F \over \p \av} = { 1 \over \Npl} \sum_i^{\Npl}{ \p \ln (g_i) \over \p \av}\, .
\end{equation} 
These $N_\alpha$ (the number of parameters in $\av$) equations generally couple to each other and require numerical solution, but see \S\ref{sec:analapp}.  \Eq{eq:alphamax} weights the shape function by detection efficiencies on the left hand side and over individual detections on the right hand side.

The basic steps in obtaining a best fit solution are as follows.  First, select a form of the PLDF to fit the data and express in the form of \Eq{eq:dfdx}.  Second, solve \Eq{eq:alphamax} for the shape parameters as just described.  Third, calculate the normalization via \Eq{eq:C} with $F$ evaluated using the best fit shape parameters.  The best fit solution is now complete and can be used  (e.g.) to estimate the NPPS as $N_{\rm tot}/N_\ast$ using \Eq{eq:Ntot}.  Fourth, estimate the errors on the fit via likelihood contours given by \Eq{eq:logL}, as described in \S\ref{sec:errors} below.

\subsubsection{Interpretation}
This formalism has a remarkably straightforward interpretation.  The normalization $C$ matches the numbers of detected and expected planets, $\Npl = \Nexp$, as seen by comparing \Eqs{eq:F}{eq:C}.    

The shape parameters similarly match the expected and detected averages of planet and host star observables.  Consider the case of powerlaw distributions with a shape function
\begin{equation} 
g = \exp\left(\av_{\rm pl}\cdot \xv + \av_\ast \cdot \zv\right)\, .
\end{equation} 
Here $\xv$ and $\zv$ are the logarithms of quantities being fit to a powerlaw, and $\av = \{ \av_{\rm pl}, \av_\ast \}$ distinguishes the powerlaw exponents for planetary and stellar parameters.
With this powerlaw shape function \Eq{eq:alphamax} gives
\begin{equation} 
\brak{\xv}_F = \brak{\xv}_{\rm obs} ~, ~\brak{\zv}_F = \brak{\zv}_{\rm obs}
\end{equation} 
where
\begin{subeqnarray}
\{ \brak{\xv}_F, \brak{\zv}_F  \}  &\equiv& {1\over F} \int \{\xv , \zv \} \eta  \mF_\ast g \dxN d\zv    \slabel{eq:Favg}\\
\{ \brak{\xv}_{\rm obs} , \brak{\zv}_{\rm obs} \}  &\equiv& {1 \over \Npl} \sum_i^{\Npl} \{ \xv_i , \zv_i \} \, .
\end{subeqnarray} 
Non-powerlaw shape functions can investigate more detailed properties of the detections than simply the average value of observables.

\subsubsection{Errors}\label{sec:errors}
To quantify the uncertainty in the shape parameters, we compare contours of the likehood $L$ to a Gaussian distribution.  The maximum likehood, $L_{\rm max}$, is given by \Eq{eq:LnoC} with the best fit $\av$.    The 1$\sigma$ uncertainties are defined by the $\ln(L) = \ln(L_{\rm max}) - 1/2$ contour.   The general $n\sigma$ contours follow  $\ln(L) = \ln(L_{\rm max}) - n^2/2$.  The uncertainty in the normalization is just the range of values taken by $C$ within the error ellipse of the $\av$.  

The 1D errors on an individual shape parameter (a component of $\av$) are usually given assuming other parameters are held at their best fit values.  When there is strong covariance, and a very elongated error ellipse (not the case in this study), 1D errors are not very meaningful. 

The errors on $\av$  describe the precision of a fit, not the quality.  A large planet sample precisely defines the average powerlaw even if the actual PLDF deviates strongly from a powerlaw.  The quality of fit can be determined by Kolmogov-Smirnoff (K-S) tests.

This derivation ignores uncertainties in the detection efficiencies and the planetary and stellar parameters. Here, we describe how to include these errors but do not include them in our analysis. Uncertainties associated with an individual detection, $i$, affect the value of the shape function, $g_i$, that appear in the likelihood analysis.  To include these uncertainties, $g_i$ should be a weighted integral over the uncertainties
\begin{equation} \label{eq:gerr}
g_i \equiv \int \phi_i(\xv' - \xv_i,\zv' - \zv_{j(i)}) g(\xv',\zv';\av) d\xv'd\zv'
\end{equation} 
where $\phi_i$ is an appropriately normalized Gaussian distribution (for instance) centered on the best fit values of the planet detection and its host star.  Ignoring uncertainties is equivalent to setting the error distributions, $\phi_i$, to $\delta$-functions.

Uncertainties in the target star properties affect the number of expected planets via the shape integral $F$.  Errors can be included similarly to \Eq{eq:gerr} as
\begin{subeqnarray} \label{eq:Ferr}
F ={1 \over N_\ast}\sum_j^{N_\ast} \int  \phi_j(\zv' - \zv_{j}) \eta(\xv,\zv')g(\xv,\zv') d\xv d\zv' \slabel{eq:Ferra} \\
\approx  \int  \phi(\zv' - \zv)\mF_\ast(\zv) \eta(\xv,\zv')g(\xv,\zv') d\xv d\zv' d\zv \, . \slabel{eq:Ferrb}
\end{subeqnarray} 
The two forms above show how errors could be applied to each star $j$, or (more approximately) how average errors could be assigned to the stellar distribution function.  The integration over both $\zv$ and $\zv'$ in \Eq{eq:Ferrb} is potentially confusing.  The integration over $\zv$ is over the stellar distribution and replaces the sum over individual stars in \Eq{eq:Ferra}, while the integral over $\zv'$ covers the range of uncertainties allowed by the error distributions $\phi$.  

Estimates of the total number of planets should also include target star uncertainties, by similarly modifying
 \Eq{eq:Ntot} (\ie setting $\eta \rightarrow 1$ in \Eq{eq:Ferr}).

Systematic uncertainties in the detection efficiencies are more difficult to quantify.  Assigning detection efficiencies are the most important (and dangerous) part of any statistical analysis.  It is safest to perform multiple analyses that cover a range of (hopefully reasonable and nearly complete) possibilities for detection efficiencies.  That approach is the one taken here.

\subsection{Which Parameters to Study}\label{sec:whichparam}
In any analysis one must choose to study some subset of planetary and stellar parameters and to ignore others.  For a robust fit, the PLDF should include all parameters that cause the detection efficiency to change significantly across the parameter space studied.   Ignoring a parameter in the PLDF amounts to averaging the PLDF over the relevant parameter.  However as expressed in \Eq{eq:Nexp}, the correct way to predict the yield of a survey is to weight the PLDF by the detection efficiencies before averaging.  

A safer way to ignore a parameter (relevant to the detection efficiency) is to first fit it and then marginalize over it.  With this caveat made explicit we now show how the general formalism treats the case of a PLDF that depends only on planetary parameters, $\xv$, or only on stellar parameters, $\zv$, (most famously stellar metallicity).

\subsubsection{Planetary Parameters Only}\label{sec:planetonly}
For a shape function, $g(\xv;\av)$, that is independent of stellar properties, the shape integral simplifies to
\begin{equation} \label{eq:Nexpsimp}
F =  \int \eta_\ast(\xv) g(\xv;\av) d\xv \, .
\end{equation} 
where the stellar-averaged efficiencies are
\begin{equation}\label{eq:etastar} 
\eta_\ast (\xv) = \int \eta(\xv, \zv) \mF_\ast(\zv) d\zv . 
\end{equation} 
The fit procedure is simplified in this case.  The stellar averaging is done a single time, and not for every choice of $\av$ during optimization.

This case is the one most relevant to this work.  Specifically in Section \ref{sec:results} we consider a shape function
\begin{equation} \label{eq:gpow}
g = \exp(\alpha x + \beta y) =  \left(R \over R_o\right)^{\alpha}\left(P \over P_o\right)^\beta\, .
\end{equation} 
This  powerlaw distribution in planetary radius and orbital period, identifies the planetary parameters as
\begin{equation} \label{eq:xv}
\xv = \{x,y\} = \{ \ln(R/R_o), \ln(P/P_o)\}
\end{equation} 
 with the help of a reference radius $R_o$ and period $P_o$, and defines the shape parameters $\av = \{\alpha, \beta\}$ as powerlaws.  Appendix \ref{sec:analapp} gives an analytic solution for these best fit powerlaw exponents for an ideal transit survey with unity discovery efficiency and no (unvetted) false positives.

\subsubsection{Stellar Parameters Only}
We now consider fitting a PLDF that depends on stellar parameters only.  As cautioned at the beginning of this section, it would be a bad idea to do this for a transit survey since the detection efficiencies always vary with orbital period, and also vary with planetary radius unless small planets with a discovery efficiency below unity are ignored.  We consider this case for other survey types and for completeness.

Since the PLDF is independent of $\xv$ by assumption, we can integrate the differential distribution of \Eq{eq:dfdx} over a volume, $V_{\xv}$, of planetary parameter space.  Defining $G \equiv g V_{\xv}$ gives
\begin{equation} 
f(\zv) = C G(\zv; \av_\ast) \, ,
\end{equation} 
which represents the NPPS with characteristics $\zv$.
The shape integral simplifies slightly to
\begin{eqnarray}
\Nexp &=& N_\ast C  \int \brak{\eta(\zv)}  \mF_\ast(\zv)  G (\zv,\alpha_\ast)d\zv\\
\brak{\eta(\zv)} &\equiv& \int \eta(\xv,\zv)\dxN/V_{\xv} \, , 
\end{eqnarray} 
by averaging the efficiencies over planetary parameters.

The best fit solutions are still given by \Eqs{eq:C}{eq:alphamax}, and $\ln g_i$ can be replaced with $\ln G_i$ (no approximation needed) to remove all references to the differential distribution.

\begin{figure*}[tb!] 
\if\submitms y
 	\includegraphics[width=3.3in]{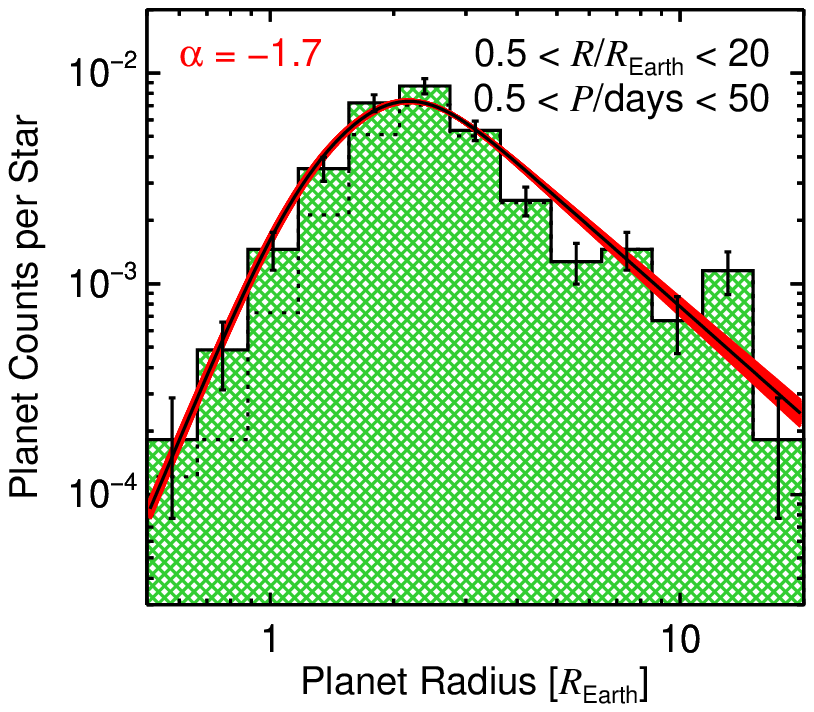}
	\includegraphics[width=3.3in]{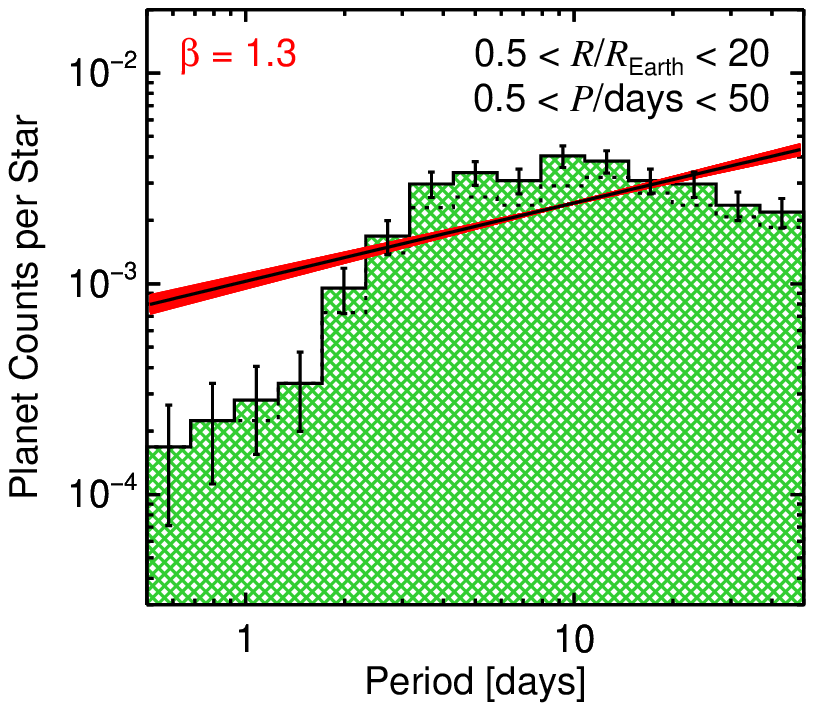}
 \else
      \includegraphics[width=3.4in]{rhistr0.5p0.5.eps} 
      \includegraphics[width=3.4in]{phistr0.5p0.5.eps}
       
\fi
   	\caption{The histograms give uncorrected \Kepler planet  candidate counts around Solar type stars.  Counts are binned by planet radius, $R$ (\emph{left}) and orbital period, $P$ (\emph{right}) and reported per star, normalized to bin width (\ie divided by $\Delta \ln R$ or $\Delta \ln P$, respectively).  Error bars measure the square root of planet counts per bin.  The main filled histogram is for detections with S/N $>$ 10, while the dotted histogram only includes detections with S/N $>$ 20 (as reported by BKep11).
 Black curves show the best fit powerlaw $\propto R^\alpha P^\beta$.  This powerlaw is convolved with the selection effects and marginalized over $P$ (or $R$) before plotting against the planet counts.   The 1$\sigma$ error ellipse of the maximum likelihood fit covers $\alpha = -1.73 \pm 0.07$ and $\beta = 1.22 \pm 0.04$ as shown by the wide red curve. 
The drop in planet counts below $\sim 2 R_\oplus$ is primarily due to selection effects, as evidenced by the  corresponding drop in the powerlaw fit when projected into observational space. 
The deficit of planets at $P \lesssim 3$ days is real and not explained by any selection effect.  Because of this short period deficit, the period distribution is poorly described by a single unbroken powerlaw.}
   	\label{fig:hists}
\end{figure*}

\begin{figure*}[tb!] 
\if\submitms y
 	\includegraphics[width=3.3in]{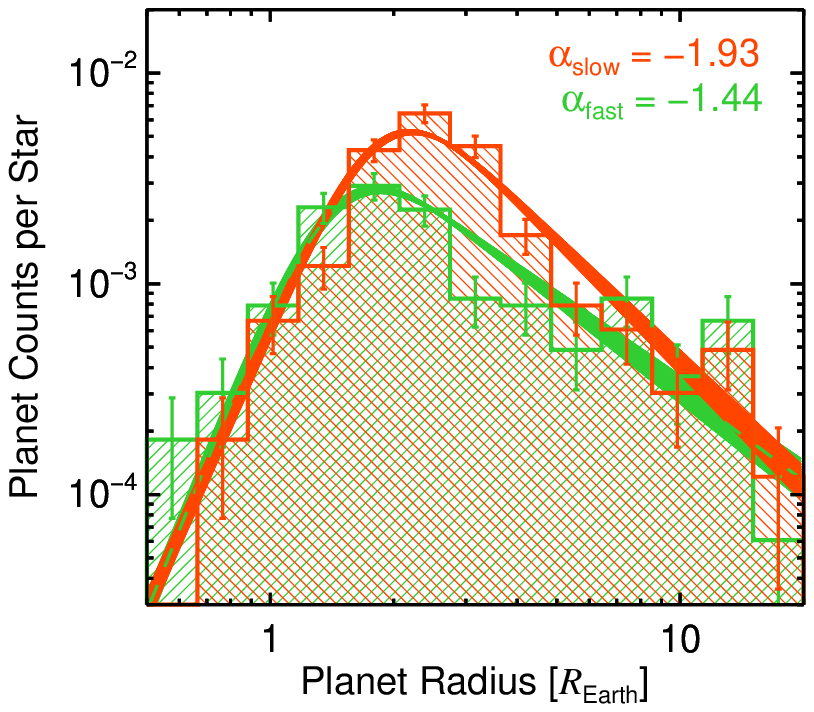}
 	\includegraphics[width=3.3in]{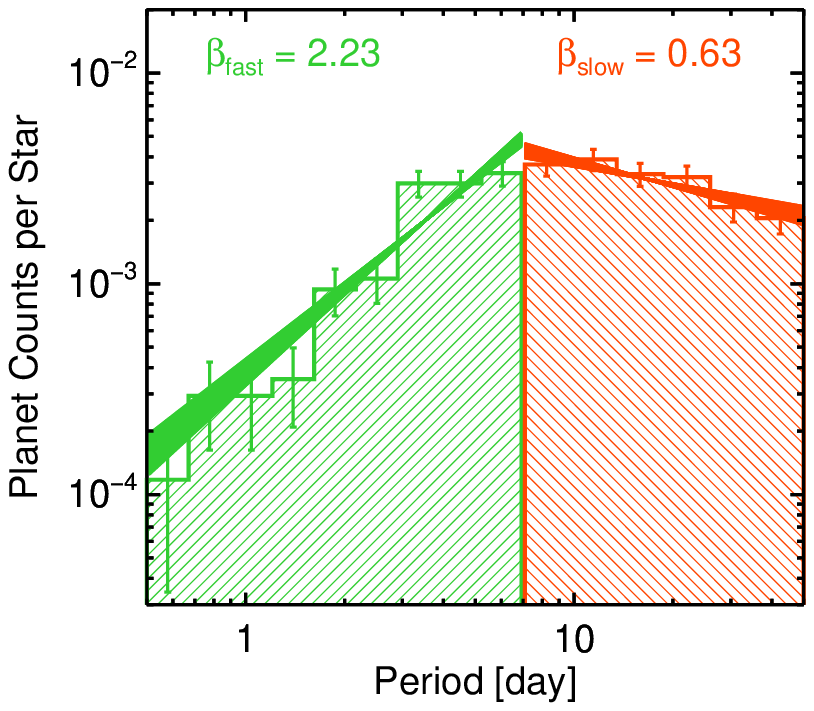}
 \else
        \includegraphics[width=3.4in]{shortlong_r.eps} 
      \includegraphics[width=3.4in]{shortlong_p.eps}
   \fi
   	\caption{Similar to \Fig{fig:hists} but the planets are divided into a short period ($P < 7$ days, in green) and a longer period ($P > 7$ days, in orange) sample.  Each sample is independently fit to a powerlaw PLDF ($\propto R^\alpha P^\beta$).  Continuity at the period boundary is \emph{not} enforced so that one sample does not affect the other.  The best fit powerlaw indices are labelled ``fast" and ``slow" for the short and long period samples, respectively.
	The  difference in the period distributions ($\beta$) has extremely high significance.  The radius powerlaw $\alpha$ is steeper for the longer period sample, a result with 2.3$\sigma$ significance.  Thus the ratio of small to big planets is higher at longer periods.}
   	\label{fig:fastslow}
\end{figure*}

\begin{figure*}[tb!] 
\if\submitms y
 	\includegraphics[width=3.3in]{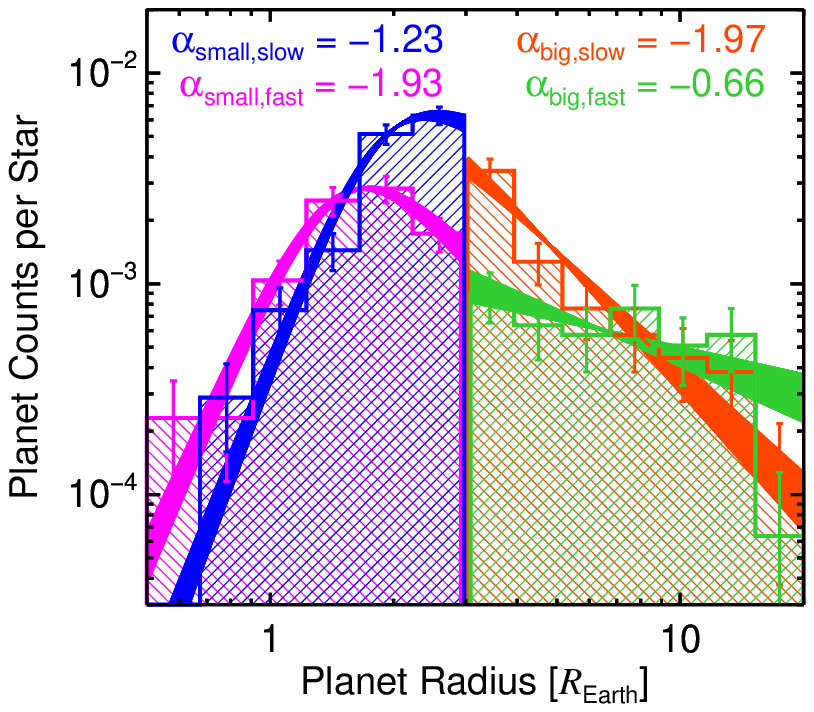}
 	\includegraphics[width=3.3in]{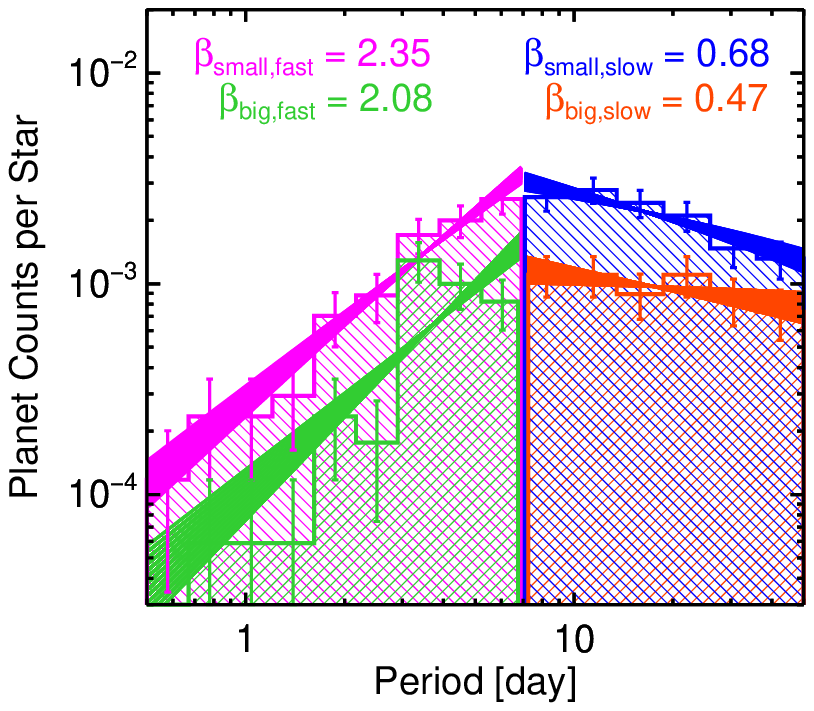}
 \else
        \includegraphics[width=3.4in]{quad_r.eps} 
      \includegraphics[width=3.4in]{quad_p.eps}
    
\fi
   	\caption{The planet sample is divided into four quadrants with separations in period at $7$ days and in radius at $3 R_\oplus$.  Each quadrant is separately fit to a powerlaw PLDF ($\propto R^\alpha P^\beta$).  The result is plotted in observational space, and the values of the best fit exponents are shown.  The differences between the subsample distributions and their significance is discussed in the text and shown in \Figs{fig:conts}{fig:dfdlnR}.}
   	\label{fig:quad}
\end{figure*}

\begin{figure}[tb] 
\if\submitms y
 	\includegraphics[width=5in]{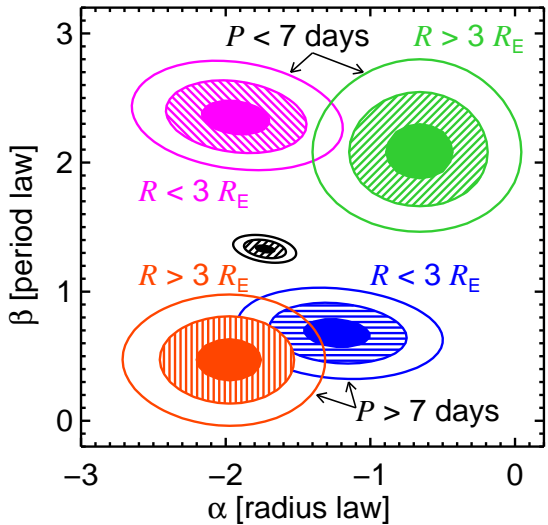}
 \else
      \includegraphics[width=3.5in]{quadcontour.eps} 
\fi
   	\caption{Error ellipses (1,2 \& 3$\sigma$) for the powerlaw fits to the PLDFs ($\propto R^\alpha P^\beta$) of different planet samples.  The black contours show the fit to the full planet sample.  Fits to the four quadrants of the planet sample shown in \Fig{fig:quad} are also given.  The difference in the period distribution of short and long period planets is highly significant.  The behavior of the  size distributions is more complex and has $\sim 2$---$3\sigma$ significance.  For shorter periods, the size distribution steepens (more negative $\alpha$) between the large and small planet samples.  Longer periods show the opposite trend --- the size distribution flattens going from bigger to smaller planets.}	 
   	\label{fig:conts}
\end{figure}

\begin{figure}[tb] 
\if\submitms y
 	\includegraphics[width=5in]{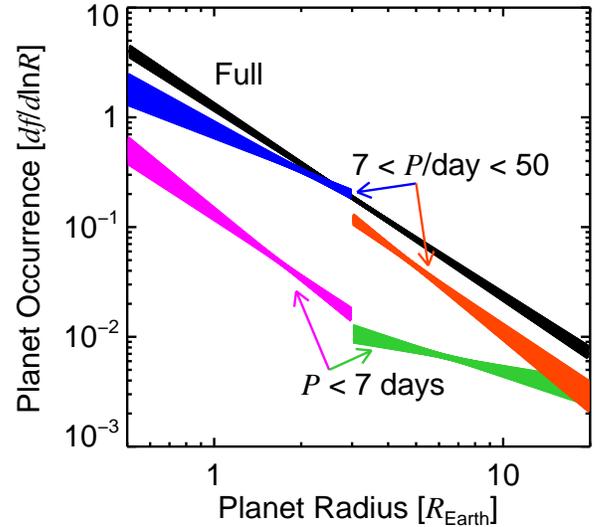}
 \else
      \includegraphics[width=3.5in]{dfdlnR.eps} 
\fi
   	\caption{Size distributions, marginalized over period.  Fits are to a powerlaw PLDF, for both our full planet sample and the four quadrants shown in \Figs{fig:quad}{fig:conts}.  Planets around $\sim 3 R_\oplus$ appear preferentially depleted at orbital periods below $\sim 7$ days.  This deficit could be due to the loss of volatiles from lower mass giant planets that approach their host stars too closely.}
   	\label{fig:dfdlnR}
\end{figure}

\section{Powerlaw Fits to {\it Kepler} Data}\label{sec:results}
We now use the \Kepler planet candidates announced by BKep11 to constrain the underlying PLDF (planetary distribution function) as a powerlaw,
\begin{equation} \label{eq:pow}
	{\p f \over \p \ln R \p \ln P}  = 
	C  \left(R \over R_o\right)^{\alpha}\left(P \over P_o\right)^\beta \, ,
\end{equation} 
in planet radius, $R$, and orbital period $P$.   \Eq{eq:pow} represents the NPPS (number of planets per star) per logarithmic interval in $R$ and $P$.
The actual PLDF can deviate from a powerlaw (and the data show that it does).  Nevertheless, powerlaws are a common and useful approximation that captures basic trends in the data.  Identifying deviations  from powerlaw behavior reveals the relevant scales of planet formation and migration, and we show that \Kepler data identify these trends extraordinarily well.

We study the full planet sample around Solar type stars in \S\ref{sec:full}.   In \S\ref{sec:shortlong}, we compare the distributions of shorter and longer period planets. Further comparing results for larger and smaller planets in \S\ref{sec:quad} shows that the sample has a deficit of intermediate-sized planets at the shortest orbital periods. For all these correlations, the S/N threshold has limited impact (\S\ref{sec:SN}).

\subsection{The Full Sample}\label{sec:full}
Our most complete sample of planets covers $0.5 < P/{\rm day} < 50$ and $0.5 < R/R_\oplus < 20$. (See \S\ref{sec:planetsample} for details.) 
\Fig{fig:hists} plots planet counts of this full sample.  The planet counts are binned by $R$ ($P$) in the left (right, respectively) panel.   Raw counts are divided by the number of stars surveyed and the logarithmic bin size.\footnote{E.g.\ divided by $\ln(R_{i+1}/R_i)$ for a bin between $R_i$ and $R_{i+1}$ and similarly for period bins.}
  Normalized this way, the data measure the PLDF as weighted by the detection efficiencies, $\eta$, 
\begin{equation} \label{eq:marg}
	\brak{\eta {d \ln f \over d \ln R}} ~{\rm and}~\brak{\eta {d \ln f \over d \ln P}}
\end{equation} 
where brackets indicate marginalization over  $P$ and $R$, respectively.    Aside from sampling noise, these values are independent of the survey size or choice of bin width. 

The best fit PLDF is ``projected" into observational space in \Fig{fig:hists}.  This projection requires weighting by the detection efficiencies of transits, $\eta_{\rm tr}$, and discovery, $\etadisc$, before marginalizing over $P$ (or $R$), as in \Eq{eq:marg}.  The comparison of the fits to histograms is somewhat misleading. The actual analysis uses unbinned data and fits $\alpha$ and $\beta$ simultaneously, not separately to the size and period distributions.  However the comparison of uncorrected counts to a projected PLDF does mimic the fitting procedure, which applies the efficiencies to the ``theory" (\ie powerlaw) not the detections.

We now explain how the powerlaw slopes are altered by the projection into observational space.   Though the best fit period distribution in the right panel of  \Fig{fig:hists} has $\beta = 1.3$, the curve in observational space has $\beta_{\rm obs} \simeq 0.3$, smaller by  about one.  A decrease of precisely  2/3 is due to the transit efficiency.  The remainder (here $\simeq -1/3$) is due to the period dependence of $\etadisc$ and depends on the amount of small planets in the sample.  From  \Eq{eq:discratio}, this correction could be as large as $-1.07$, the period exponent of $\etadisc$ at low efficiencies.  Though not easily visible, there is modest curvature to the projected period powerlaw due to lower discovery efficiencies at longer periods.

The discovery efficiency has a much more dramatic effect on the projected size distribution, whose curve is evident in the left panel of \Fig{fig:hists}.   There are no free parameters to adjust either the position or angle of the break, because $\etadisc$ is fixed.   Above $\sim 3 R_\oplus$, the \Kepler discovery efficiency is unity for all periods considered.  Thus the slope of the projected size distribution at large radii matches the bestfit $\alpha$.   The break starts at $R \lesssim 3 R_\oplus$, because $\etadisc$ drops below $90\%$ for these sizes, as shown in \Fig{fig:discrat}.  At small $R$, the powerlaw slope of the projected radial PLDF is $\alpha + 6.34$, as fixed by the radial dependence of  the discovery efficiency in \Eq{eq:discratio}.

The thickness of the fits shown in \Fig{fig:hists} (and other figures) indicate $1\sigma$ deviations from maximum likelihood.  \Fig{fig:conts} shows the corresponding error ellipse in the $\alpha$ and $\beta$ parameters (the filled black oval is for the full planet sample).  The 1$\sigma$ errors on $\alpha$ and $\beta$ reported in \Tab{tab:fits} are one-dimensional errors holding the other parameter fixed.  As explained in Section \ref{sec:method}, the normalization $C$ is not independently varied during the fit procedure, but follows from the values of $\alpha$, $\beta$ and the planet-to-star ratio.  The errors on $C$ indicate the range of values taken inside the 1$\sigma$ error ellipse of $\alpha$ and $\beta$.

Our best fit size distribution has $\alpha = -1.73 \pm 0.07$.  This value indicates that smaller planets are much more abundant, but that larger planets contain most of the mass.  At constant planet density, $\alpha < -3$ would give small planets most of the mass.  However larger planets tend to have lower density, $\rho$.  Take $\rho \propto R^{-0.7}$ as an example.\footnote{Appropriate if 1 $R_\oplus$ planets have $\rho \approx 5 \gcc$ and 10 $R_\oplus$ planets have $\rho \approx 1 \gcc$.}  Then $\alpha < -2.3$ would give small planets more of the mass (over the range where the density law holds).  The conclusion that larger planets dominate the mass distribution holds even if the giants are quite inflated.

While planet counts drop towards small sizes, so does the projection of the bestfit powerlaw, as described above.  The good qualitative agreement between the counts and the curve (in the left panel of \Fig{fig:hists}) means that the drop in planet counts is mostly due to selection effects.  The actual size distribution continues to rise towards smaller planets.

The best fit period law, $\beta = 1.33 \pm 0.03$, indicates that planets are more closely packed further from the star, for logarithmic intervals in $P$ or semimajor axis $a$.   For $\beta > 3/2$ the planet density per linear interval in $a$ would increase.

However \Fig{fig:hists} shows that a single powerlaw is a qualitatively poor fit to the data due to a sharp drop in planet counts below $P \sim 3$ days.   Planet counts flatten at longer periods.  As explained above, this means that the actual period distribution continues to rise towards longer periods.  This break in the period distribution motivates our division of the planet sample below.

\subsection{Short Vs. Long Periods}\label{sec:shortlong}
Since the period distribution deviates strongly from a simple powelaw, we divide the planets into short and long period samples.  We use $P = 7$ days as the dividing line.  The choice gives us comparable numbers of planets in each sample and also gives an adequate range of periods over which to measure a powerlaw slope.  Planet counts in different cuts are summarized in the $N_{\rm pl}$ column of  \Tab{tab:fits}.

\Fig{fig:fastslow} shows the planet counts and the results of independent powerlaw fits to the short and long period data.  These fits are not broken powerlaws, as the values are not required to match at the period boundary.  Two powerlaws vastly improve the qualitative fit to the data.  More complicated functions could more precisely describe the location and nature of the period break, but are not considered here.

The radius distribution in the left panel includes overlapping histograms for the short and long period samples.  The longer period (``slow") sample has a steeper radius distribution than the shorter period (``fast").  The difference of $\alpha_{\rm slow} - \alpha_{\rm fast} \approx 0.5 \pm 0.2$ has about 2.5$\sigma$ significance.  By eye, the difference in the distribution is clearly driven by the flatness of the fast sample between $\sim 3$---$15 R_\oplus$.  To clarify this behavior we now examine the differences between small and large radius  planet samples.

\subsection{Quadrants}\label{sec:quad}
We subdivide the planets into four samples  by splitting both the  shorter and longer period populations (still defined relative to $7$ days) into ``small" and ``big"  samples relative to $3 R_\oplus$.  The results of the four independent powerlaw fits are plotted against planet counts in \Fig{fig:quad}.   \Fig{fig:conts} plots error ellipses to compare the slope changes and their statistical significance.  The error ellipses of the subsamples are much larger than the full sample.  Smaller number counts and the decreased leverage of the reduced range in $R$ and $P$ (over which to define a slope) both play a role.

The period distributions are only modestly affected by this split.  At both short and long periods, the best fit $\beta$ of the small and big samples agree within the statistical uncertainties.  Nevertheless some qualitatively interesting features appear in the right panel of \Fig{fig:quad}.  The small planet counts begin their decline below about 7 days (as noted by HKep11).  The large planet sample remains flat towards shorter periods --- with evidence of a peak at $P \approx 3$ days.  This peak is responsible for the flat appearance of the overall period distribution down to 3 days, as seen in the right panels of \Figs{fig:hists}{fig:fastslow}.

The behavior of the size distributions is even more intriguing.  For big planets,  the best fit radius law  changes by $\alpha_{\rm big,slow} - \alpha_{\rm big,fast} = -1.31 \pm 0.47$, with almost $3\sigma$ significance.  Since the discovery efficiency of \Kepler is near unity for these large planets, no obvious systematic uncertainties should be at work.  For small planets, the change in the size distributions, $\alpha_{\rm small,slow} - \alpha_{\rm small,fast} = 0.70 \pm 0.47$.  While of  more modest significance, this change is of the opposite sign.

\Fig{fig:dfdlnR} plots the best fit size distributions for each quadrant,  with the full planet sample shown for comparison.  At longer periods the size distribution flattens towards smaller radii, \ie $\alpha$ is less negative for the small planets.
For the shortest period planets the behavior is precisely opposite.  The size distribution is quite shallow at large sizes and steepens for the small radius sample. 

The implications of this behavior  are intriguing.  The processes that form and/or migrate planets inside $\sim 7$ days clearly disfavors planets of a few to several $R_\oplus$.  A plausible scenario (but lacking in specifics) follows.  At short periods, only the most massive giants --- true Jovians with $R \gtrsim 11 R_\oplus$ can efficiently retain their atmospheres.  Lesser giants are stripped of their atmospheres, including ices that sublimate from the surface.  The demotion of ice and gas giants to small rocky cores can simultaneously explain the flatter size distribution at large sizes and the steeper distribution at small sizes. 

If this hypothesis is born out by further observations and theoretical study, it would provide yet another pillar of support for the core-accretion hypothesis and for migration as the source of short period planets. 

\subsection{Systematic Uncertainty}\label{sec:SN}
To test the quality of the fits, we consider an increased S/N threshold.
Increasing this threshold preferentially removes smaller planets from the sample. In principle the appropriate S/N threshold is set by the analysis of the discovery efficiency, $\etadisc$.  As discussed in \S\ref{sec:planetsample}, increasing the S/N threshold is a conservative way to check how the assumed $\etadisc$ impacts the results.

\Fig{fig:hists} includes dotted histograms that correspond to the higher S/N $> 20$ sample.  Planet counts are noticeably lower below $3 R_\oplus$ and across all periods.

\Tab{tab:fits} includes fits to the higher S/N planet samples.  All of the qualitative trends discussed in this section persist in the higher S/N sample and with only mostly reduced significance.   Indeed, the interesting trends we identified for small planets was a steepening of the size distribution at small periods and a flattening at longer periods.  It would be rather difficult for systematic uncertainties to conspire to produce both trends.  

Despite this robustness, $\etadisc$ is crucial for the statistical analysis of small planets.  The final rows of  \Tab{tab:fits} shows that setting $\etadisc = 1$ (but still including the transit probability) gives very discrepant results.  These discrepancies are expected due to the low discovery efficiency below $2 R_\oplus$, as shown in  \Fig{fig:discrat}.

\section{Planet Occurrence}\label{sec:npps}
To derive the number of planets per star, NPPS, we consider our non-parametric analysis.  For $R > 2.0 R_\oplus$ and $P< 50$ days, we find $f_{2,50} = 0.19$ planets per star, consistent with HKep11.

When we consider smaller planets the NPPS is roughly one.  For $R > 0.5 R_\oplus$, S/N thresholds of 10, 15 and 20 give  $f_{0.5,50} = 1.36$, 1.05 and 0.72 planets per star.   Since the S/N = 20 threshold is conservative, it seems likely $f_{0.5,50} > 1$.  

The likely existence of more than one planet per Solar type star, especially within 50 days, is remarkable.   However it does not imply that the Sun is exceptional for lacking a planet so close.  For $R > 1.5 R_\oplus$ there are $\sim$2---3 planets per planetary system \citep{lr11}.   Multiplicity rates will almost certainly increase when planets down to $0.5 R_\oplus$ are considered.

It is inherently speculative to extrapolate into unobserved regions of parameter space.  Nevertheless, all the powerlaw fits of the previous section have distributions that rise with period.     To emphasize the implications of this rise, we calculate the extrapolated density of Earth-sized planets at 1 AU.  The final column of  \Tab{tab:fits} expresses the normalization to the fits as 
\begin{equation} 
C_{\oplus, {\rm yr}} \equiv {\p^2f \over \p \ln R \p \ln P}(R_\oplus,{\rm year})\, ,
\end{equation} 
the NPPS in a logarithmic interval centered on an Earth radius and a year.

The most relevant fit is to the small planet, long period sample, which gives $C_{\oplus, {\rm yr}} = 2.75 \pm 0.33$.  Integrating this powerlaw gives on average $f_{1,365} \approx 3$ Earth-like planets with periods below a year.  If the distribution function of planets does not turn over shortly beyond periods of 50 days, the implications for habitable Earths around Sun-like stars is truly staggering.

\section{Concluding Discussion}\label{sec:disc}
We develop a general technique for the analysis of exoplanet survey data.  The merits of the approach, a generalization of the TT02 maximum likelihood analysis,  include the following:
\begin{itemize}
\item Data are analyzed without binning in order to preserve the statistical significance of each detection. 
\item Describing the number of planets per star (NPPS)  allows multi-planet systems to be included.
\item Selection effects are used to calculate the number of expected detections.  They are not used to enhance (or diminish) the number of actual detections, which affects error estimates. 
\item The overall normalization is analytically removed from the likelihood function.  With one less parameter in the numerical optimization, the fit is simpler and possibly more robust.  
\item Different surveys can be jointly analyzed, as shown by TT02.  Even different survey methods that probe different regions of parameter space can be combined if selection effects are well characterized.
\end{itemize}

We apply the technique to \Kepler planet candidates released by BKep11.  We focus on planets orbiting Solar type stars for which HKep11 has quantified the \Kepler discovery efficiency.  Fits to these efficiences (see \Fig{fig:discrat}) reveal a remarkably smooth variation with planet radius and orbital period that agrees with analytic predictions for transit surveys. This regular behavior gives us confidence to analyze detections down to $0.5 R_\oplus$. 

From this analysis, the best estimate of the number of planets per star bigger than $0.5 R_\oplus$ and with periods below $50 \days$ is $f  \simeq 0.7$ --- $1.4$.  Uncertainty in the discovery efficiencies dominate these estimates. 
Since planet occurrence rises towards both small sizes and longer periods, the only question is when (not if) we can claim with certainty that there is more than one planet per star.

The shape of the radius and period distributions is even more informative than absolute number counts.   The most notable trend is a difference between the size distributions of shorter and longer period planets  (below and above 7 days), plotted in \Fig{fig:dfdlnR}.  While the shorter period planets are less abundant overall, their relative deficit is most pronounced around  $\sim 3 R_\oplus$.
We interpret this finding as arising from the preferential evaparation of ice and lower mass gas giants that migrated too close to their host star.  In this picture, the steepness of the size distribution of small planets at short periods arises from the remnant cores of these stripped giants.

If this interpretation holds, it would add support to the core accretion hypothesis \citep{pollack96}, though this theory is not seriously challenged at masses below $M_{\rm Jup}$ and possibly much higher \citep{kmy10}.  In particular, cores would have to be present by the end of migration.  This constraint applies to models where the cores sediment over time \citep{hs08}.

We do not here attempt to constrain the mode of migration with these size trends.  Possibilities include smooth inward migration through the disk, a theory that was highly developed even before the discovery of extrasolar planets  \citep{lp86,ward86,pawel93}.   Planet-planet scattering \citep{cfmr08} and Kozai oscillations \citep{wm03} have also been proposed as mechaisms to deliver planets close to their hosts.  Recently \citet{wl10}, proposed a new mechanism --- secular chaos --- and compared all the proposed mechanisms to observations of hot Jupiters.  These theories and others should now be compared to \Kepler data for low mass planets as well.


\acknowledgements
Scott Kenyon's advice and encouragement made this paper possible.  I  thank Kevin Schlaufman for making Tables 1 and 2 of \citet{borucki_pc11} available in ascii format at {\wrapletters{\tt http://www.ucolick.org/$\sim$kcs/research/planets/kepler\_table.dat}}.   I am very grateful to Elizabeth Adams, Jean-Michel Dessert, Andrew Howard,  Matt Holman and Jonathan Irwin for insightful discussions and to all responsible for maintaining the venue of these conversations: the daily SSP coffee at the CfA.  This project was supported by the {\it NASA} {\it Astrophysics Theory Program} and  {\it Origins of Solar Systems Program}  through grant NNX10AF35G and by endowment funds of the Smithsonian Institute.\\

\appendix

\section{Analytic Solutions for Powerlaw Fits}\label{sec:analapp}
This appendix derives an analytic solution to the best fit powerlaws of the planetary radius and orbital period.  The transit probability is included, but all other detection efficiencies are set to unity, as appropriate for a perfect transit survey.  Since numerical fits are more flexible, this derivation is intended to provide insight and to provide a check on numerical solutions (when only the transit probability is included as a selection effect).

Consider the detection of $\Npl$ planets in a transit survey.  We fit the PLDF to the powerlaw of  \Eq{eq:gpow} (equivalent to \Eq{eq:pow} used in our main analysis of \Kepler data).  The logarithms of radius and period are defined as the parameters $x$ and $y$ as in \Eq{eq:xv}.    We define the reference planet radius, $R_o$, and orbital period $P_o$ so the (here rectangular) domain of the survey is centered on $x = y = 0$ and all detections (indexed by $i$) fall within
\begin{equation} 
-x_{\rm m} \leq x_i \leq x_{\rm m}~~;~~ -y_{\rm m} \leq y_i \leq y_{\rm m}\, .
\end{equation} 

The transit efficiency of \Eq{eq:etatr} can be written as
\begin{eqnarray} 
\eta_{\rm tr} = \eta_{\rm tr,o}  \exp \left(-{2y\over 3}\right) ~;~\eta_{\rm tr,o} \equiv  \left(3 \pi\over  G \rho_\ast  P_o^{2}  \right)^{1/3} \, ,
\end{eqnarray} 
and we ignore other selection effects.

The number of expected transits for such a perfect survey is $\Nexp = N_\ast C F_{\rm perf}$ with
\begin{equation} 
F_{\rm perf} = 4 \eta_{\rm tr, o}{ \sinh(\alpha x_{\rm m}) \over \alpha}{\sinh[(\beta - 2/3 )y_{\rm m}] \over \beta - 2/3}\, .
\end{equation} 
given by \Eq{eq:Nexpsimp} (which averages the general \Eq{eq:Nexp} over stellar parameters).

The best-fitting $\alpha$ and $\beta$ follow from \Eq{eq:alphamax} as
\begin{equation} 
\label{eq:analsol}  
	 \coth(\alpha x_{\rm m}) - {1 \over \alpha x_{\rm m}}  = {\brak{x}_{\rm obs} \over x_{\rm m}} ~~;~~
	\coth[(\beta - 2/3) y_{\rm m}] - {1 \over(\beta - 2/3) y_{\rm m}}  = {\brak{y}_{\rm obs} \over y_{\rm m} }\, . 
\end{equation}  
where the obsevational mean $\brak{x}_{\rm obs} \equiv \Npl^{-1}\sum_ix_i$, and similarly for $\brak{y}_{\rm obs}$.  The relevant transcendental function, $T(\gamma) = \coth(\gamma) - 1/\gamma$ is monotonic in $\gamma$ with $T \rightarrow \pm 1$ as $\gamma \rightarrow \pm \infty$ and $T(\gamma) \approx \gamma/3$ for $\left|\gamma\right| \ll 1$.
The closed form solutions of \Eq{eq:analsol} have several features that relate to more complete analyses:
\begin{itemize}
\item The powerlaws increase monotonically with the average value of the (log of the) observables, independent of how the values are distributed around that mean.  Higher order shape functions can describe more detailed features in the data.
\item The period powerlaw $\beta$ in the PLDF is $2/3$ larger than the period law describing the detections.  The transit efficiency, $\eta_{\rm tr}$ causes this increase.  In general any powerlaw selection effect adjusts from the observed to the underlying distribution this way.
\item The overall magnitude of the detection efficiency does not affect powerlaws, or shape parameters in general.  For instance, uncertainty in $\eta_{\rm tr,o}$ due to uncertainty in stellar densities does not affect the shape function.  The overall planet occurrence of course does depend on the magnitude of detection efficiencies, via the normalization given by \Eq{eq:C}. 
\item Solutions for the best fit powerlaws decouple.  In general, the solution for two shape parameters decouple if the shape parameters (and the physical parameters they describe) are separable in the shape function $g$ \emph{and} the relevant physical parameters are separable in the detection efficiencies.   This separability condition is not met for out full treatment of the \Kepler data.  While the efficiency ratio $r_{\rm disc}$ of \Eq{eq:discratio} is separable, the efficiency itself, $\etadisc = r_{\rm disc}/(1+r_{\rm disc})$, is not.  This leads to the modest covariance between $\alpha$ and $\beta$ seen in \Fig{fig:conts}.
\end{itemize}

\if\bibinc n
\bibliography{refs}
\fi

\if\bibinc y

\fi

\begin{deluxetable}{rrrcccccr}  

\tablecolumns{9}
\tablecaption{Joint Powerlaw Fits to \Kepler Planet Candidates in ``Solar" Subsample}
\small
\tablehead{   
 \colhead{} & \multicolumn{3}{c}{\underline{Planet Sample}}\vline &  \multicolumn{5}{c}{\underline{Fit Properties}} 
\\ 
  \colhead{Tag} 	 &
  \colhead{Radii} &
  \colhead{Periods} 	 &
   \colhead{$N_{\rm pl}$}\tablenotemark{b} \vline & 
     \colhead{$N_{\rm tot}$} & 
  \colhead{planets\tablenotemark{c}} & 
  \colhead{$\alpha$} 	          &
  \colhead{$\beta$} & 
 \colhead{$C_{\oplus,{\rm yr}}$} 
\\
\colhead{} &
 \colhead{[$R_\oplus$]\tablenotemark{a}} &
  \colhead{[days]}&
\colhead{} \vline & 
  \colhead{(corrected)} &
   \colhead{per star} &
  \colhead{(radius law)} &
  \colhead{(period law)} & 
\colhead{}
}
\startdata
Full & 0.5 --- 20 & 0.5 --- 50 & 562 & $1.4 \times 10^{5}$ & 2.38 & $-1.73 \pm  0.07$ & $1.33 \pm  0.03$ & $23.40 \pm 0.50$ \\
Full/$R2$ & 2 --- 20 & 0.5 --- 50 & 372 & $1.2 \times 10^{4}$ & 0.21 & $-1.88 \pm  0.11$ & $1.22 \pm  0.04$ & $20.30 \pm 1.35$ \\
\hline
\sidehead{(Short vs. Long Period: ``Fast" vs.\ ``Slow")}
Fast & 0.5 --- 20 & 0.5 --- 7 & 211 & $1.0 \times 10^{4}$ & 0.18 & $-1.44 \pm  0.11$ & $2.23 \pm  0.12$ & $(15 \pm 1)$E2 \\
Slow & 0.5 --- 20 & 7 --- 50 & 351 & $1.3 \times 10^{5}$ & 2.26 & $-1.93 \pm  0.10$ & $0.63 \pm  0.10$ & $3.57 \pm 0.06$ \\
Fast/$R2$ & 2 --- 20 & 0.5 --- 7 & 111 & $1.3 \times 10^{3}$ & 0.02 & $-1.09 \pm  0.17$ & $2.27 \pm  0.18$ & $(9.8 \pm 1.3)$E2 \\
Slow/$R2$ & 2 --- 20 & 7 --- 50 & 261 & $9.7 \times 10^{3}$ & 0.17 & $-2.31 \pm  0.15$ & $0.54 \pm  0.11$ & $4.71 \pm 0.47$ \\
\hline
Small & 0.5 --- 3 & 0.5 --- 50 & 389 & $1.3 \times 10^{5}$ & 2.30 & $-1.52 \pm  0.16$ & $1.43 \pm  0.04$ & $32.05 \pm 2.16$ \\
Big & 3 --- 20 & 0.5 --- 50 & 173 & $4.5 \times 10^{3}$ & 0.08 & $-1.42 \pm  0.16$ & $1.08 \pm  0.06$ & $5.26 \pm 0.32$ \\
\hline
\sidehead{(Quadrants)}
Small/Fast & 0.5 --- 3 & 0.5 --- 7 & 148 & $1.5 \times 10^{4}$ & 0.25 & $-1.93 \pm  0.24$ & $2.35 \pm  0.14$ & $(33 \pm 3)$E2 \\
Big/Fast & 3 --- 20 & 0.5 --- 7 & 63 & $6.8 \times 10^{2}$ & 0.01 & $-0.66 \pm  0.24$ & $2.08 \pm  0.22$ & $(1.7 \pm 0.3$)E2 \\
Small/Slow & 0.5 --- 3 & 7 --- 50 & 241 & $7.6 \times 10^{4}$ & 1.31 & $-1.23 \pm  0.23$ & $0.68 \pm  0.12$ & $2.75 \pm 0.33$ \\
Big/Slow & 3 --- 20 & 7 --- 50 & 110 & $3.4 \times 10^{3}$ & 0.06 & $-1.97 \pm  0.23$ & $0.47 \pm  0.17$ & $2.09 \pm 0.23$ \\
\hline
\sidehead{(S/N$>$20)}
Full/sn20 & 0.5 --- 20 & 0.5 --- 50 & 453 & $8.0 \times 10^{4}$ & 1.37 & $-1.44 \pm  0.07$ & $1.30 \pm  0.04$ & $12.54 \pm 0.31$ \\
SmFa/sn20 & 0.5 --- 3 & 0.5 --- 7 & 105 & $6.5 \times 10^{3}$ & 0.11 & $-1.34 \pm  0.30$ & $2.16 \pm  0.16$ & $(7.3 \pm 0.8)$E2 \\
BiFa/sn20 & 3 --- 20 & 0.5 --- 7 & 62 & $6.7 \times 10^{2}$ & 0.01 & $-0.51 \pm  0.25$ & $2.07 \pm  0.22$ & $(1.3 \pm 0.2)$E2 \\
SmSl/sn20 & 0.5 --- 3 & 7 --- 50 & 178 & $3.6 \times 10^{4}$ & 0.62 & $-0.63 \pm  0.30$ & $0.77 \pm  0.14$ & $1.73 \pm 0.28$ \\
BiSl/sn20 & 3 --- 20 & 7 --- 50 & 108 & $3.4 \times 10^{3}$ & 0.06 & $-1.88 \pm  0.24$ & $0.50 \pm  0.17$ & $1.98 \pm 0.22$ \\
\hline
\sidehead{(TRANSIT PROBABLILITY ONLY,  Perfect Discovery Efficiency, $\etadisc = 1$)  }
Full/tp & 0.5 --- 20 & 0.5 --- 50 & 562 & $1.4 \times 10^{4}$ & 0.24 & $-0.18 \pm  0.04$ & $1.04 \pm  0.03$ & $0.64 \pm 0.01$ \\
SmFa/tp & 0.5 --- 3 & 0.5 --- 7 & 148 & $1.5 \times 10^{3}$ & 0.03 & $1.19 \pm  0.18$ & $1.94 \pm  0.14$ & $41.49 \pm 3.59$ \\
BiFa/tp & 3 --- 20 & 0.5 --- 7 & 63 & $6.8 \times 10^{2}$ & 0.01 & $-0.52 \pm  0.25$ & $2.08 \pm  0.22$ & $(1.4 \pm 0.2)$E2 \\
SmSl/tp & 0.5 --- 3 & 7 --- 50 & 241 & $7.0 \times 10^{3}$ & 0.12 & $2.45 \pm  0.18$ & $0.26 \pm  0.12$ & $0.02 \pm 0.00$ \\
BiSl/tp  & 3 --- 20 & 7 --- 50 & 110 & $3.4 \times 10^{3}$ & 0.06 & $-1.89 \pm  0.24$ & $0.48 \pm  0.17$ & $1.89 \pm 0.21$ 
\enddata
\tablenotetext{a}{Planet radii are computed form the products of $R/R_\ast$ and $R_\ast$ given in Tables 1 and 2 of BKep11.}  
\tablenotetext{b}{$N_{\rm pl}$ :Number of \Kepler planet candidates with S/N $>$ 10 (or 20 where  indicated) as reported in BKep11 .}
\tablenotetext{c}{See \S\ref{sec:npps} for more accurate non-parametric fits to the planet occurrence.}
\label{tab:fits}
\end{deluxetable}

\end{document}